# Step-edge epitaxy for borophene growth on insulators

Qiyuan Ruan, Luqing Wang, Ksenia V. Bets, and Boris I. Yakobson*

Department of Materials Science and NanoEngineering, and Department of Chemistry, Rice University, Houston, Texas 77005, United States

**ABSTRACT**:

Borophene – a monoatomic layer of boron atoms – stands out among two-dimensional (2D) materials, with its versatile properties of polymorphism, metallicity, plasmonics, superconductivity, tantalizing for physics exploration and the next-generation devices. Yet its phases are all synthesized on and stay bound to metal substrates, hampering both characterization and use. The growth on the inert insulator would allow post-synthesis exfoliation of borophene, but its weak adhesion to such substrate results in a very high 2D-nucleation barrier preventing clean borophene growth. This challenge can be circumvented in a devised and demonstrated here, with *ab initio* calculations, strategy. Naturally present 1D-defects, the step-edges on h-BN substrate surface, enable boron epitaxial assembly, reduce the nucleation dimensionality and lower the barrier by an order of magnitude (to 1.1 eV or less), yielding $v_{1/9}$ phase. Weak borophene adhesion to the insulator makes it readily accessible for comprehensive property tests or transfer into the device setting.

**Introduction**

The early investigation of the small boron clusters demonstrated the stability of the quasi-planar configurations.[1–4] Later, the prediction of the fullerene-like $B_{80}$ with its peculiar at the time structure motivated the concept of the hexagonal hollows (HH) in planar boron configurations.[5] Soon, the computational exploration of the two-dimensional (2D) boron sheets formed by a combination of triangular lattice and various concentrations and distribution of HH resulted in the introduction of α, β, and γ phases of borophene.[6,7] Using metal-substrates, the borophene forms synthesized by now include $v_{1/5}$ and $v_{1/6}$ on Ag(111),[8–12] honeycomb $v_{1/3}$ on Al(111),[13,14] $v_{1/12}$ on Au(111),[15] a novel $v_{1/5}$ phase on Cu(111),[16] $v_{1/6} - \chi_6$ on Ir(111).[17]

The abundance of unique properties in borophene — nontrivial band structure with Dirac cones[18] and nodal lines,[19] 2D-plasmonics,[20] and BCS superconductivity[21] — makes it one of the most promising synthetic 2D materials.[22–24] However, its strong adhesion to metallic substrates[25–27] that initially enabled successful syntheses of the 2D instead of naturally-favored 3D phase becomes a challenge to material separation for analysis and applications. This motivates alternative approaches,[28] yet the reproducible borophene, free or on insulator, remains elusive. The weaker interaction with the insulator substrate would allow the lift-off,[29] however, it might not provide sufficient stabilization and preference towards the 2D growth. Here we explore the possibility to employ further dimensionality reduction to 1D during nucleation that should enable subsequent borophene growth on less-adhesive substrates. The naturally present one-dimensional defects within substrate material such as surface steps would provide a convenient nucleation site with increased strength on interaction.[30–32] The interaction with steps is particularly important for growth on van der Waals crystal where substrate terraces are fully saturated while step edges may expose dangling bonds. The vicinal plane (0001) of hexagonal boron nitride (h-BN) presents an opportune inert insulator substrate material with weakly binding continuous planes while the step-edges display the high affinity to boron atoms as well as satisfactory lattice matching with borophene lattice, as we quantify below.

**Results and Discussion**

Stability and energetics of the borophene phases on various substrates are strongly connected to the balance between the triangular lattice regions (electron donors) and hexagonal holes (electron acceptors) in addition to the electron exchange with the substrate.[6] For example, the Ag(111) substrate favors the formation of the $v_{1/5}$ and $v_{1/6}$ phases, as experiments show, while in $v_{2/15}$, $v_{1/8}$, $v_{1/9}$ phases are more stable in a vacuum.[33] Among them, $v_{1/9}$ phase or $\alpha$-borophene, has the lowest energy in vacuum due to the atomic arrangement that satisfied the self-donors/acceptors condition precisely.[6] Therefore, even though the charge donation from the substrate plays an important role for borophene growth in some conditions, on metals or electrides,[29] it is not always necessary for borophene to accept electrons in its formation.

As a wide bandgap inert insulator, h-BN is expected to provide minimal electron interaction with the borophene layer. Confirming this, we find that the adhesion of the five major phases of borophene to the underlying h-BN is virtually identical and on par with the van der Waals attraction in bilayer graphene or h-BN sheets, while weaker than the borophene-Ag interaction (the latter three values agree with previous

works[34–36]), **Fig. 1a**. Furthermore, the Bader analysis shows negligible charge transfer (0.0002 $e^-$ per boron atom) from the h-BN substrate to $v_{1/9}$ borophene compared to the transfer from Ag(111) to $v_{1/6}$ phase (0.03 $e^-$ per boron atom). The Fermi level of borophene is located well within the bandgap of h-BN (**Fig. S1**), suppressing the charge transfer between the two. Low interaction with the substrate suggests that the stability of borophene phases on the h-BN would be similar to that in a vacuum, which our direct energy calculations confirm (**Fig. 1a**). Therefore, the $v_{1/9}$ is chosen as a representative phase to explore the growth on the h-BN substrate.

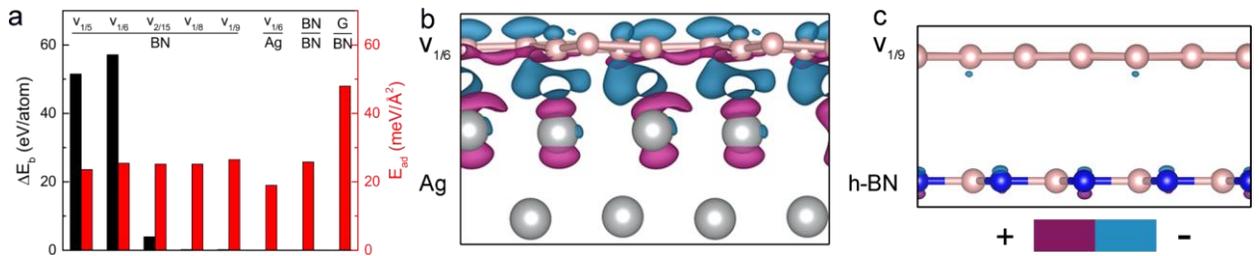

**Figure 1**. (a) The adhesion energy of borophene phases to h-BN or Ag(111) substrates, $E_{ad} = (E_B + E_{sub} - E_{B@sub})/area$, compared to that in bilayer AB-stacked graphene (G) or h-BN, all red-shaded wide bars and right axis. The binding energy for five common phases, relative to $v_{1/9}$, $\Delta E_b = (E_{B@BN} - E_{v_{1/9}@BN})/N_B$, all on h-BN, all-black narrow bars, refer to the left axis. Calculated charge density difference of (b) borophene $v_{1/6}$ on Ag(111) with isosurface at 0.0016 $e^-/\text{Å}^3$ and (c) $v_{1/9}$ on h-BN with isosurface at 0.0008 $e^-/\text{Å}^3$, show negligible electron donation from the h-BN, compared with metal substrates.

To assess the feasibility of the on-the-step 1D nucleation of borophene, we follow the nanoreactor diagram,[37,38] where the building blocks are sequentially attached to the growing nucleus, and the resulting energies are computed and compared. To determine the dominant mechanism, we consider nucleation both at the substrate step and on the terrace away from it (**Fig. 2a**). For the evaluation of the nucleus docked to the step, we choose h-BN zigzag edge due to good lattice match with borophene $v_{1/9}$ resulting in a minimal misfit and consequently the lowest interfacial energy (**Fig. 2b**). As discussed below, the B-terminated edge indeed leads to minimal lattice mismatch, however, nucleation near N-terminated edge exhibits drastically different growth sequence leading to the borophene orientation with significant lattice mismatch and ultimately rising energy (see **Fig.S5** for details). This observation led us to consider growth near the B-terminated h-BN edge as the primary nucleation route for the 1D case. An ultrahigh vacuum experimental condition or annealing may help avoid the possible passivation of pristin B-terminated edge.

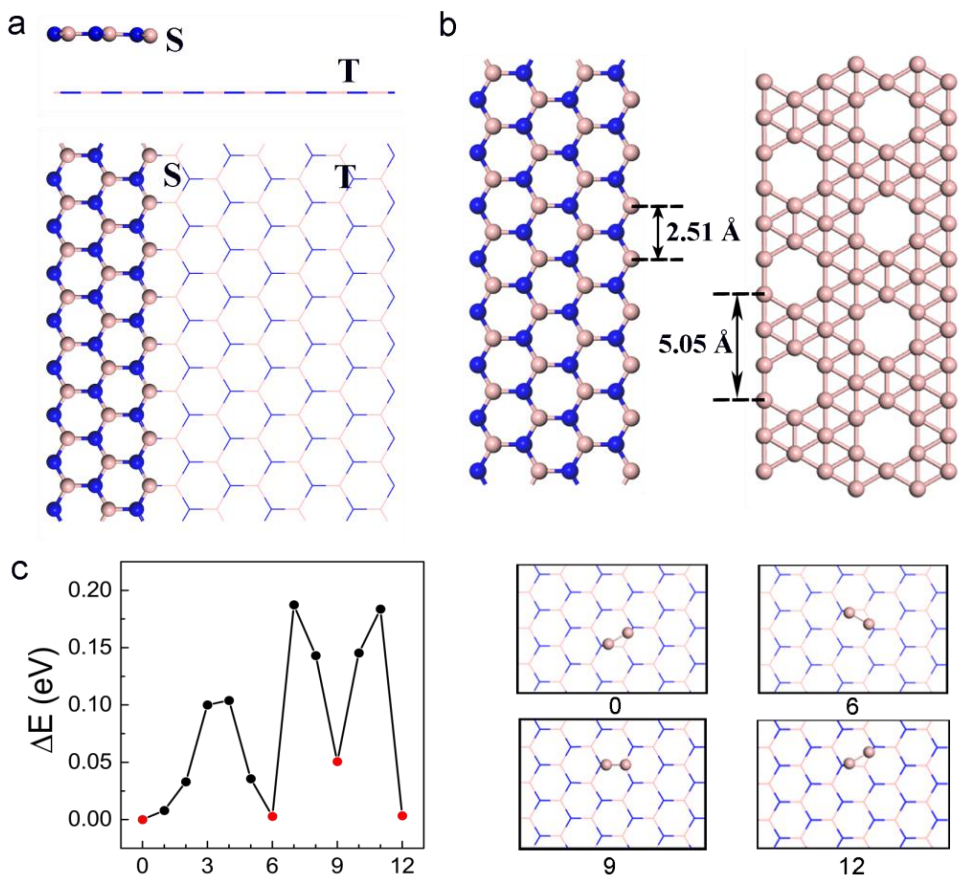

**Figure 2**. (a) Side and top view of the AA'-stacked bilayer h-BN substrate with step (S) and terrace (T) sites. (b) The h-BN and borophene $v_{1/9}$ lattices; best-match periods marked by the dashed lines. (c) The computed barriers for $B_2$ diffusion on the h-BN substrate. The shown global minima and metastable configurations correspond to the red dots on the energy plot.

The implementation of the nanoreactor model requires the choice of the building blocks-units for the growing crystal. Assuming that the adsorption of the source material in the course of the molecular beam epitaxy (MBE) happens on the entire surface of the substrate, we evaluate the stability of boron monomers and dimers. We find that on the h-BN a dimer $B_2$ has energy 0.94 eV lower than two monomers, and hence $B_2$ is chosen as an accretion-unit in the nanoreactor model. We also confirm dimers' significant mobility on the surface of the h-BN by calculating the diffusion barriers (**Fig. 2c**), equal 0.1 eV for the migration within one hexagon of the h-BN lattice (**Fig. 2c**, 1-6), and 0.2 eV across the hexagons (**Fig. 2c**, 6-12). For comparison, boron monomers show a diffusion barrier of 0.3 eV. The low barrier values indicate significant mobility of the $B_2$ on the substrate surface, further confirming its likely prevalence as a building unit, an immediate growth precursor. It is worth pointing out that boron dimer is not a necessary building block of borophene growth, boron monomer may still lead to the same calculation results.

Following the nanoreactor procedure, at each step, we consider all possible dimer attachment sites and evaluate their energies. Configuration with the lowest energy is chosen as the "winner" and propagated to the next growth step, towards the $v_{1/9}$ formation. We characterize the accretion process by tracking the Gibbs free energy change (**Fig. 3a**):

$$G(n) = E_{B@sub} - E_{sub} - n\,\mu_0,$$

where $E_{B@sub}$ is the total energy of the boron cluster with $n$ atoms on the h-BN substrate, $E_{sub}$ is its total energy, $\mu_0$ is the chemical potential of the boron source. For sheer reference, we first choose $\mu_0 = \mu_{v_{1/9}} = -6.46$ eV/atom, corresponding to the product-phase.

We begin by probing conventional borophene nucleation away from any steps,[25,29] on the h-BN terrace sites, statistically more abundant by orders of magnitude than any linear sites, grain boundaries or step edges. We trace the nucleus extension from a single dimer up to the $B_{44}$ cluster, selected configurations shown in **Fig. 3b**. Throughout the process, the island-nucleus maintains a compact truncated polygonal shape allowing for a larger bulk-to-edge ratio ($\sim\sqrt{n}$) among borophene atoms, and mostly consists of the triangular lattice regions; occasional pentagonal defects occur but "heal" and vanish at the next step. The first hollow hexagon, characteristic for borophene structures, appears later in trajectory, at $n = 38$, in **Fig. 3b** right. With our reference choice of low $\mu_0$, at the level of the bulk $v_{1/9}$ phase (black line in **Fig. 3a**), the 2D island excess energy per atom would approach zero with the larger size, but the perimeter adds a contribution $\sim\sqrt{n}$, asymptotically. Therefore, the G-curve in the 2D case keeps rising with no display of a maximum; no finite barrier appears at this $\mu_0$, while the maximum Gibbs free energy 15.8 eV we reach at $B_{40}$. Generally, the energies for all sizes $n$ significantly exceed those on Ag(111), e.g., by 4-5 eV for the $B_{20}$ cluster,[25] indicating that vdW interaction with the h-BN substrate is indeed too weak to enable the nucleation on the flat terrace. As one may notice, our lowest energy path does not necessarily include the lowest energy clusters at each *n*-step but is rather the lowest energy-increment sequence of structures. For example, at *n* = 32, 34, and 36 lower energy boron clusters exist (see **Fig. S3**) but did not appear in the trajectory as they could not be obtained by adding a dimer to the one-prior island structure. Among these three, noteworthy is the $B_{36}$, experimentally obtained and mentioned as the possible borophene precursor.[39] Interestingly, the largest observed in our sequence nucleus does not follow $v_{1/9}$ lattice pattern, indicating that terrace nucleation would be characterized by the presence of structural defects or perhaps the mixing of different phases.

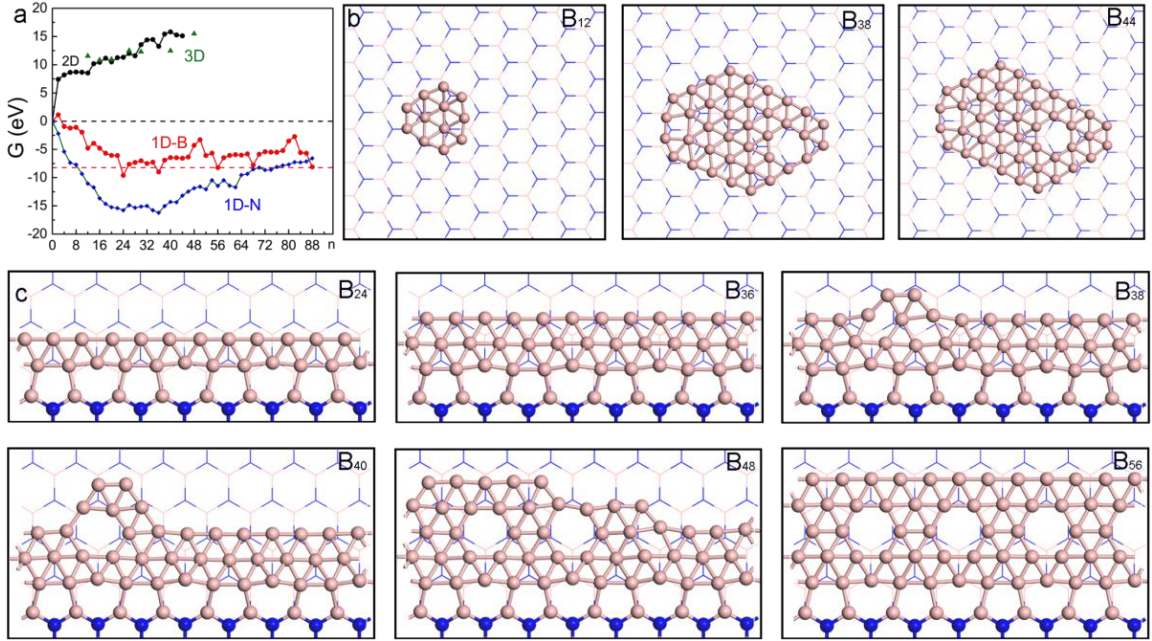

**Figure 3**. (a) Change of the Gibbs free energy versus number *n* of B atoms during borophene islands nucleation, on the terrace (2D, black squares) and at the B- or N-terminated step edges, red 1D-B or blue 1D-N. The energies for several 3D clusters (**Fig. S6**) on the h-BN terrace are plotted for comparison (green triangles). Sample configurations during nucleation on the terrace (b) and next to step (c), with *n* values labeled.

In drastic contrast to the terrace case, an h-BN step, either of the random-roughness origin or created by a vicinal cut, offers drastically different nucleation tendencies. As the analysis below shows, the strong covalent affinity between the substrate step and boron atoms makes growth along the step (1D) significantly lower in energy than a conventional on-terrace. As a result, borophene aggregate first extends in a linear fashion along the substrate step, completely covering h-BN with a double and then triple row of boron atoms (**Fig. 3c**, $B_{24}$, and $B_{36}$). Notably, this initial stage is followed by the formation of a characteristic $v_{1/9}$ pattern of hollow hexagons. Further growth is realized through the repetition of $B_{24}$ - $B_{56}$ segments or alike: generally, for $v_{1/9}$ phase, such segment would span 4s of B atoms, on h-BN step-base of s lattice units (**Fig. S4**). Selected configurations are shown in **Fig. 3c**, for s = 8 case. Interestingly, while the initiation of each new row is accompanied by a moderate energy barrier ~1.5 eV (see $B_2$, $B_{26}$, $B_{38}$ on **Fig. 3a**), there is no critical nucleus size, and the nucleation macro-barrier is completely bypassed through the interaction with the h-BN edge. This is a classic 1D growth behavior,[40] and also similar to that recently shown for 1D nucleation in peptides.[41] This confirms that the 1D nucleation mechanism can allow the growth of borophene on the insulating substrates, whose low areal interaction must also facilitate the post-growth separation.

In the case of 1D clusters, the presence of the edge is combined with the h-BN/borophene interface that significantly lowers the energy of the system. As a result, in simulations with our chosen $\mu_0$, the energy of the 1D cluster follows near a horizontal line (red dashed line in **Fig. 3a**) at the negative value equal to the length of the interface (or the length of the periodic cell), times the double-edge energy of the $v_{1/9}$ phase, less its binding to the h-BN step.

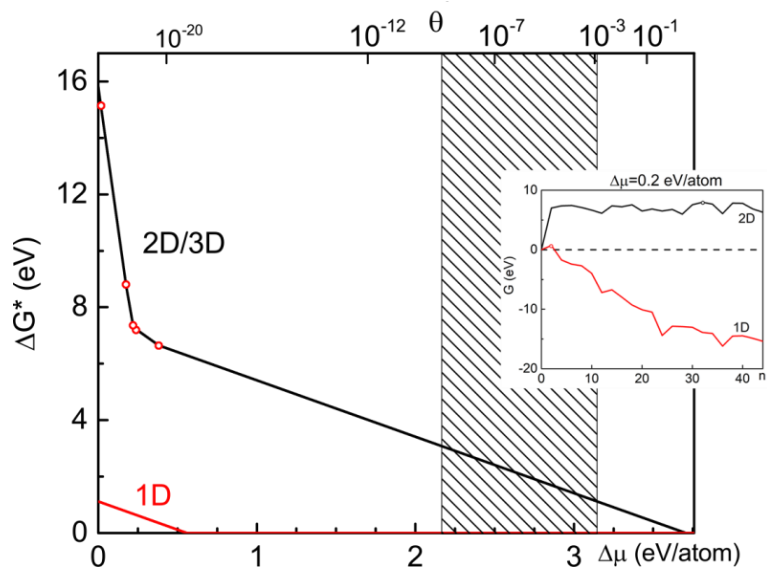

**Figure 4**. The nucleation barrier $\Delta G^*$ versus feedstock chemical potential $\Delta\mu$, ranging from the $v_{1/9}$ phase ($\Delta\mu$ = 0) to $B_2$ dimer on the h-BN surface ($\Delta\mu$ = 3.76 eV), for 2D and 1D growth (black and red lines). The shaded area marks the conditions where 2D and 3D nucleation is suppressed while 1D→2D growth is barrierless and fast, optimal for the experimental realization. Right inset: an example of the $\Delta G(n)$ curves at low $\Delta\mu$ = 0.2 eV corresponding to the critical nucleus at $n$ = 32 and 2, for 2D and 1D, respectively. Left insets: the critical nuclei schematics at various $\Delta\mu$, their $\Delta G^*$ values marked by small red circles.

The energy maximum, usually associated with the nucleation barrier, depends on the chemical potential of $B_2$. While the choice $\mu_0 = \mu_{v_{1/9}}$ (product borophene-phase) is a convenient reference for plotting G($n$), growth requires the boron feedstock on the substrate to be higher in energy than the product, by $\Delta\mu = \mu_0 - \mu_B \geq 0$, defined by experimental conditions. Various $\Delta\mu$ do alter the barrier height $\Delta G^*$ in a way very different for 1D and 2D paths, as **Fig. 4** shows.

For 1D growth, the barriers vanish at $\Delta\mu$ > 0.5 eV (negligible $B_2$ surface concentration, according to **Fig. S9**), while the 2D nucleation barrier is prohibitively high, ~ 6 eV. The latter is reduced gradually, for richer $B_2$ feedstock, to also vanish at $\Delta\mu$ > 3.5 eV (interestingly, it corresponds to the substrate nearly fully covered with $B_2$, fusing into a borophene phase with no need of nucleation). However, while permitting 2D-nucleation, higher $\Delta\mu$ would also allow unwanted 3D-clusters (of energies similar to the 2D islands, Fig. 3a). This suggests that high-quality synthesis should be sought at moderate $\Delta\mu$ values, lower surface $B_2$ coverage, when both 3D and 2D-on-terrace nucleation is suppressed. Assuming even a relatively low portion of step sites on h-BN surface (~$10^{-3}$), the difference between 3D/2D and 1D-initiation barriers of 1.1 eV or more (at $\Delta\mu$ < 3.15 eV) is sufficient to ensure at least $10^4$ purity of 1D-nucleation into planar borophene, at T = 800 K. On the other hand, we note that the corresponding coverage $\theta \sim 10^{-3}$ ($B_2$ concentrations $10^{11}$ cm$^{-2}$) is sufficient to support the borophene growth at speeds up to $10^4$ μm/h, in a ballpark with realistic rates so far. The borophene growth feasibility range (a shaded area in **Fig. 4**) is also bounded on the left by kinetic requirement: the $B_2$ diffusion flux $\sim\theta\sqrt{(D/\tau)}$ should be sufficient to support visible growth speed of ~75 Å/h or more, so $\theta$ should not be too small (here D is diffusion coefficient of $B_2$ and $\tau$ is its residence time on the substrate, see **SI.6**).

The notion of 1D step-edges promoting borophene growth can be generalized on other van der Waals layered substrates. Graphite naturally comes to one's mind, with the edge of most popular graphene, of a structure and even lattice constant similar to h-BN, and therefore to borophene. Can graphene be a candidate for borophene growth? One prerequisite is that the substrate should allow unobstructed motion of building units-molecules ($B_2$) across the terrace towards the step. To elucidate the situation, we examined the swap energy of a boron atom or dimer in graphene, to verify whether building block molecules are prone to embedding within the graphene terrace as well as for the h-BN terrace. The swap energy, i.e., the energy difference for the substitution of carbon atoms by boron atom or dimer, is calculated by $E_{swap} = E_{final} - E_{initial}$, where $E_{initial}$ and $E_{final}$ are the total energies before and after substitution, respectively. The swap energies of boron atom and dimer in graphene are 0.25 eV and -0.75 eV, while those in h-BN are 0.68 eV and 3.80 eV. The small and even negative swap energies in graphene suggest that building block molecules commonly would be trapped on the terrace, instead of moving towards graphene steps and nucleating the borophene phase, as was experimentally observed.[12] At the same time, the large swap energies prevent the h-BN terrace from trapping building block molecules. All substitution structures are shown in **Figs. S7** and **S8**. This demonstrates the unsuitability of graphene as a substrate for borophene growth, suggesting that similar analyses should be performed for all potential candidates.

**Conclusions**

The analysis above maps a route for 2D borophene synthesis on non-metallic weakly-bonding substrates. This can be achieved through growth dimensionality reduction, realized by the boron interaction with the active step-edges on the vicinal surface of the otherwise inert h-BN crystal. Negligible contact charge transfer and weak interaction with h-BN planes make borophene energetics very similar to that in a vacuum, with $v_{1/9}$ phase ($\alpha$-borophene) most likely to form. Its most probable growth precursor is $B_2$ dimers, with low-barrier (~0.2 eV) surface diffusion and no trapping by the h-BN lattice (in contrast to a graphene substrate, where boron atoms get easily swapped with carbon and trapped, as we show).

We determine that not far from equilibrium, the barrier for borophene nucleation on a pristine h-BN plane is very large, >15 eV. It can only be suppressed at highly nonequilibrium conditions (Δμ ≳ 3.5 eV), when borophene, however, would form concurrently with various 3D-clusters, together yielding amorphous films.

At the same time, 1D growth at the h-BN step essentially bypasses the nucleation barrier. Boron atoms have a high affinity to the h-BN step, and its B-terminated zigzag edge, in particular, shows a good epitaxial lattice-match with the emergent $v_{1/9}$ phase. Together, these factors result in a barrier for sequentially initiating atomic rows as low as ~1.1 eV, or even vanishing further away from equilibrium when sufficient surface concentration of the $B_2$ precursor is reached. The two conditions of (i) boron supply sufficiently high to support growth rate kinetically, yet (ii) low enough to avoid formation of 3D and amorphous forms, together determine the optimal range (**Fig. 4**), important to guide experimentation. The use of the steps on the surface of the 2D vdW substrate to facilitate borophene synthesis is not limited to the h-BN, although careful consideration of the substrate specifics is essential. Needless to say, successful realization of the step-edge assisted borophene growth on inert insulating substrate would significantly advance its research in two ways. A number of physical properties, from plasmons to superconductivity, can be probed directly on insulating substrates, not obscured by the metallic conductivity of the latter. Furthermore, distinctly weak adhesion should allow facile exfoliation and transfer for unambiguous microscopy characterization or potential device applications.

**Methods**

DFT calculations were performed using DFT-D2 van der Waals (vdW) approach[42] in conjunction with the Perdew–Burke–Ernzerhof (PBE) generalized gradient approximation[43] in the framework of the all-electron projector augmented wave (PAW) method[44] as implemented in the Vienna Ab-initio Simulation Package.[45] The plane-wave kinetic energy cutoff was 500 eV. To eliminate the interactions between a boron sheet or cluster and its periodic images, a vacuum slab of 15 Å has been included.